\documentclass[aps,pra,a4paper,twocolumn,showpacs,showkeys,floatfix,superscriptaddress]{revtex4-1}

\usepackage{graphicx}
\usepackage{epstopdf}
\usepackage[dvips]{color}
\usepackage{amsmath}
\usepackage{multirow}

\graphicspath{{images/}}

\newcommand{\figwidth}{0.98\columnwidth}

\newcommand{\dimpy}{\textnormal{({C}$_{7}$H$_{10}$N)$_{2}$Cu$_{(1-x)}$Zn$_{x}$Br$_{4}$~}}
\begin{document}
	\title{Electron spin resonance study of spin relaxation in the strong-leg spin ladder with nonmagnetic dilution}

	\author{Yu. V. Krasnikova}
	\email{Electronic address: krasnikova.mipt@gmail.com}
	
	\affiliation{P. L. Kapitza Institute for Physical Problems RAS, Kosygin
		Str. 2, 119334 Moscow, Russia}
	
	\affiliation{International Laboratory for Condensed Matter Physics, National Research University Higher School of Economics, Myasnitskaya Str. 20, 101000 Moscow, Russia}

	\author{V. N. Glazkov}
	\affiliation{P. L. Kapitza Institute for Physical Problems RAS, Kosygin
		Str. 2, 119334 Moscow, Russia}
	
	\affiliation{International Laboratory for Condensed Matter Physics, National Research University Higher School of Economics, Myasnitskaya Str. 20, 101000 Moscow, Russia}
	
	\author{A. Ponomaryov}
	\affiliation{Dresden High Magnetic Field Laboratory (HLD-EMFL), Helmholtz-Zentrum Dresden-Rossendorf, D-01328, Dresden, Germany}
	
	\author{S. A. Zvyagin}
	\affiliation{Dresden High Magnetic Field Laboratory (HLD-EMFL), Helmholtz-Zentrum Dresden-Rossendorf, D-01328, Dresden, Germany}
	
	\author{K. Yu. Povarov}
	\affiliation{Laboratory for
		Solid State Physics, ETH Z\"{u}rich, 8093 Z\"{u}rich, Switzerland}
	
	\author{S. Galeski}
	\altaffiliation[Present address: ]{Max-Planck-Institute for Chemical Physics of Solids, N\"{o}thnitzer Stra{\ss}e 40, 01187 Dresden, Germany}

	\affiliation{Laboratory for Solid State Physics, ETH Z\"{u}rich,
		8093 Z\"{u}rich, Switzerland}
	
	\author{A. Zheludev}
	\affiliation{Laboratory for Solid State Physics, ETH
		Z\"{u}rich, 8093 Z\"{u}rich, Switzerland}
	
	\date{\today}
	
	\begin{abstract}
		We have studied electron spin resonance (ESR) absorption spectra for the non-magnetically diluted strong-leg spin ladder magnet ({C}$_{7}$H$_{10}$N)$_{2}$Cu$_{(1-x)}$Zn$_{x}$Br$_{4}$ (abbreviated as DIMPY) down to 450 mK. Formation of the clusters with non-zero net magnetization is confirmed; the cluster-cluster interaction is evidenced by the concentration dependence of ESR absorption. High-temperature spin-relaxation time was found to increase with non-magnetic dilution. The ESR linewidth analysis proves that the Dzyaloshinskii-Moriya (DM) interaction remains the dominant spin-relaxation channel in diluted DIMPY. Experimental data indicate that the dilution results in the weakening of the effective DM interaction, which can be interpreted as total suppression of DM interaction in the close vicinity of impurity atom.
		
	\end{abstract}
	
	
	
	%75.10.Kt    Quantum spin liquids, valence bond phases and related phenomena
	%76.30.-v Electron paramagnetic resonance and relaxation
	
	\maketitle
	
	\section{Introduction.}
	Low-dimensional magnetic systems draw attention because of their rich physics and the possibility to affect their spin state and excitations spectrum \cite{ward}. Spin ladder is the model magnetic system actively studied in recent decades. This interest is governed by recent studies, confirming, e.g., field-induced transition into the Tomonaga-Luttinger liquid state \cite{ruegg, hong, povarov}, magnon condensation \cite{tsvelik, zapf, giamarchi} and superconductivity in Cu-oxide ladders \cite{uehara, sigrist, dagotto}.
	
	A simple two-leg spin ladder consists of two Heisenberg spin chains which are connected to each other by the exchange couplings. These chains form the ``legs'' of the ladder and inter-chain couplings form the ``rungs'' of the ladder. If the exchange coupling along the rung of the ladder is nonzero, the excitation spectrum has a gap between the exited states and the ground state \cite{schmidiger}. The spin ladder does not order down to $T=0$ and demonstrates quantum paramagnetic behavior. Applied magnetic field closes the gap, which results in the formation of the Tomonaga-Luttinger liquid phase \cite{ruegg, hong, povarov}. The weak inter-ladder coupling could lead to the formation of the antiferromagnetic ordering at the sufficiently low temperature under the applied magnetic field \cite{white, chaboussant}.
	
	In this paper, we present an experimental study of spin-relaxation processes in the non-magnetically diluted strong-leg spin ladder {\dimpy} (called DIMPY for short). This magnet has ladder structure and represents a unique case of the ladder with dominating coupling along the legs of the ladder \cite{shapiro, schmidiger0}. Consistency of description of DIMPY by the spin ladder model was proved by the neutron scattering measurements, the values of exchange couplings and the gap were determined from the inelastic neutron scattering experiment: $J_{leg}$=1.42 meV, $J_{rung}$=0.82 meV and $\Delta$=0.33 meV \cite{schmidiger1}. Besides of the Heisenberg exchange coupling in pure DIMPY $(x=0 \%)$ ESR spectroscopy revealed the presence of the uniform Dzyaloshinskii-Moriya (DM) interaction $D$=0.03 meV \cite{glazkov0, ozerov}.
	
	One of the ways to affect the magnetic properties of spin ladders is a non-magnetic dilution. This dilution can be performed through substitution of a magnetic ion by a non-magnetic ion (site dilution) or through substitution of one of the surrounding ions mediating the superexchange path (bond dilution). In general, impurities introduction in low-dimensional magnets could lead to uncommon effects such as induced antiferromagnetic ordering or to the formation of new objects near impurity \cite{sigrist, sandvik, poilblanc, yasuda, wessel, fukuyama, glazkov1}. Local antiferromagnetic correlations in 1D and 2D systems are known to be enhanced in the vicinity of vacancy \cite{martins}, leading to the formation of the antiferromagnetically correlated multi-spin clusters.
	
	Diluted spin ladders behave differently compared to spin chains. Non-magnetic site dilution breaks the spin chain apart, while the ladder system remains connected. Thus, the diluted spin ladder contains randomly distributed clusters of correlated spins and does not split into the independent finite size fragments. These clusters interact with each other via the effective exchange coupling mediated by the spin ladder matrix. The effective exchange coupling constant exponentially decreases with the inter-cluster distance, thus a strong-leg spin ladder (which demonstrates large correlation length along the ladder) is more suitable for the study of the cluster-cluster interaction \cite{mikeska}. Therefore, the Zn-diluted strong-leg spin ladder \dimpy{} is a perfect system to study the effects of dilution on the low-energy spin dynamics and to search for the manifestations of the cluster-cluster interactions.

	We have studied ESR response of the Zn-diluted DIMPY from 0.45 K to 300 K. At low temperatures the ESR response is dominated by multi-spin clusters. Analysis of the ESR intensity and linewidth concentration dependencies provides evidences of cluster-cluster interaction. At higher temperatures the collective gapped excitations dominate the magnetic response of the Zn-diluted DIMPY. However the dilution affects the spin-relaxation even in the high-temperature regime: the ESR linewidth decreases with increasing Zn content. Analysis of angular dependencies of the ESR linewidth for different Zn concentrations and temperatures proves that the DM interaction remains the dominant spin relaxation channel, but its effective strength is reduced by the dilution.

	\section{Samples and experimental details.\label{sect:sym}}
	
	Single crystals of the Zn-diluted DIMPY are from the same batch as the samples studied in Ref. [\onlinecite{schmidiger2}]. The crystal structure was confirmed by the x-ray diffraction using a BRUKER APEX II diffractometer, the crystals have the same space group $P2(1)/n$ as the pure compound, and the lattice parameters are close to those of pure DIMPY \cite{shapiro}, as grown crystals have a well-developed plane normal to $b$ axis and are elongated along the $a$ axis.
	
	The unit cell of DIMPY includes four copper ions which form rungs of two parallel ladders running along the $a$ axis \cite{shapiro}. Thus, there are two equivalent spin ladders, which are differently oriented with respect to the crystal. Due to the $g$-factor anisotropy, ESR responses of different ladders can be resolved, maximal separation of the absorption signals was observed for $H||(\bf{X}+\bf{Y})$, here $\bf{X}$, $\bf{Y}$ are the unit vectors parallel to the $a$ and $b$ directions, correspondingly \cite{glazkov0}.
	
	We have carried out ESR experiments for several Zn concentrations: 0\%, 1\%, 2\%, 4\%, 6\%. These are the nominal concentrations as determined by the amount of Zn salts in the growth solution. We have not checked the real concentration of zinc by a chemical analysis, however, successful modeling of the magnetization in Ref. [\onlinecite{schmidiger2}] has shown that the nominal concentration is very close to the real.
	
	The main ESR experiments were carried out in Kapitza Institute for Physical Problems with the home-made ESR spectrometers equipped with helium-4 and helium-3 cryostats. The lowest temperature was 450 mK, the magnetic field up to 10 T was supplied by a superconducting cryomagnet, and the microwave frequencies from 17 to 40 GHz were supplied by a set of microwave generators.
	
	Additionally, high sensitivity X-band (9.3 GHz) ESR experiments were performed in HLD-EMFL in Dresden at BRUKER ELEXYS E500 spectrometer equipped with the helium-4 flow cryostat and the automatic goniometer.
	
	\begin{figure}[h]
		\includegraphics[width=\figwidth]{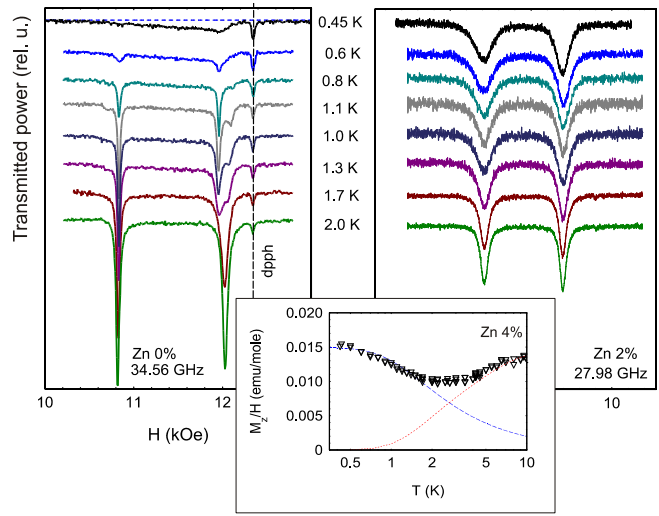}
		\caption{Temperature dependence ($T$=0.45 K -- 2 K) of ESR absorption in DIMPY: left panel --  pure DIMPY, $f$=34.56 GHz, right panel -- 2 \% Zn diluted sample, $f$=27.98 GHz. Sample orientation corresponds to maximum ESR component splitting [$H$$\parallel$({\bf{X}+\bf{Y}})]. Dashed vertical line on the left panel shows position of DPPH marker ($g$=$2.00$). Insert: Temperature dependence of scaled ESR absorption intensity for 4 \% Zn diluted DIMPY. Blue dashed line -- model of paramagnetic centers, red dotted line -- model of 1D spin gap magnet.}\label{fig:lines}
	\end{figure}
	
	\begin{figure}[h]
		\includegraphics[width=\figwidth]{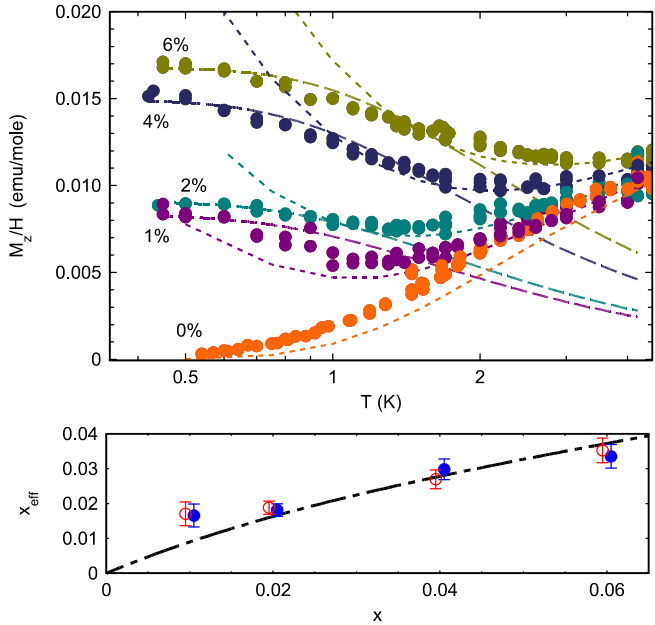}
		\caption{Upper panel: Temperature dependence of ESR absorption intensity ($T$=0.45 K...4.2 K at $f$=29.7 GHz) for the left component of ESR line in Zn diluted DIMPY for all studied concentrations of Zn (0\%, 1\%, 2\%, 4\%, 6\%). Sample orientation corresponds to maximum of ESR component splitting [$H$$\parallel$({\bf{X}+\bf{Y}})]. ESR intensity is scaled to 4 K static susceptibility data of Ref. [\onlinecite{schmidiger2}]. Different symbols correspond to listed Zn concentrations. The green dashed lines are static susceptibility curves calculated within the ``spin islands'' model \cite{schmidiger2}. The brown dashed lines are fits of low-temperature (below 700 mK) ESR intensity in the model of free paramagnetic $S=1/2$ centers $I$=$A\cdot\tanh{(g{\mu}_{B}H/(2T))}$. Lower panel: apparent concentration of paramagnetic centers $x_{eff} \propto A$ for different nominal Zn concentrations $x$, dashed-dotted line is model curve  $x_{eff}=\frac{x}{2}[{1+(1-x)^{23}}]$ (see text).}\label{fig:intens}
	\end{figure}

	\begin{figure}[h]
		\includegraphics[width=\figwidth]{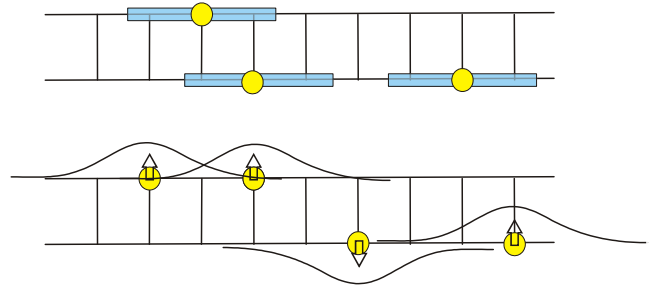}
		\caption{Schematic representation of diluted ladders. Circles mark impurity ion positions. Upper panel: Model for triplet-dominated regime, rectangles show area of DM interaction suppression around the impurity ion. Lower panel: Modeling of ESR intensity and ESR linewidth in impurity-dominated regime, envelopes of spin clusters are shown by curves, arrows mark direction of the net magnetization of cluster. }\label{fig:ladders}
	\end{figure}
	
	\begin{figure}[h]
		\includegraphics[width=\figwidth]{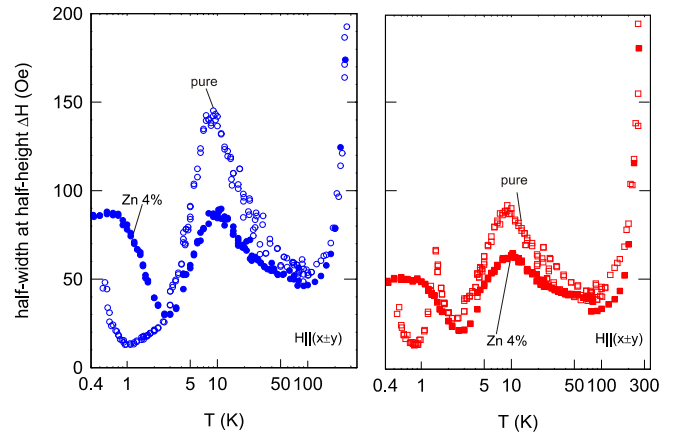}
		\caption{Comparison of linewidth temperature dependencies ($T$=0.45 K...300 K) for both components of ESR absorption in pure and 4 \% Zn diluted DIMPY. Left panel -- Left component of ESR absorption, right panel -- right component. Open symbols -- Data for pure DIMPY, filled symbols -- for 4 \% Zn diluted sample. Sample orientation corresponds to maximum of ESR component splitting ($H$$\parallel$({\bf{X}+\bf{Y}})). Full temperature range was covered by several experimental setups operating on different microwave frequencies: for pure sample $f\approx $17 GHz at $T=2...25$K and $77...300$K and $f \approx $33 GHz at $T=0.45...3$K and $25...80$K, for Zn 4 \% diluted sample $f \approx$17 GHz at $T=3...77...300$K and $f \approx $28 GHz at $T=0.45...4$K}\label{fig:widetemp}
	\end{figure}
	
	\begin{figure}[h]
		\includegraphics[width=\figwidth]{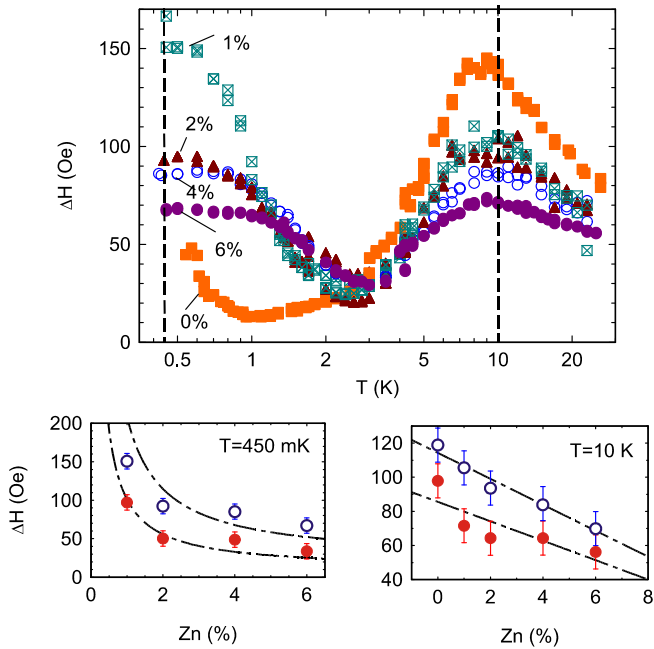}
		\caption{Upper panel: Linewidth temperature dependence for the left component of ESR absorption for all studied concentrations of Zn (0\%, 1\%, 2\%, 4\%, 6\%) in DIMPY. Data at $T$=0.45 K...4.2 K are obtained at $f$=29.7 GHz, data at 4.2 K...25 K are obtained at $f$=17.4 GHz. Different symbols correspond to listed Zn concentrations. Vertical dashed lines show temperatures used to plot lower panels graphs. Lower panels: Linewidth dependence on concentration for both components of ESR absorption line (blue open symbols -- left component, red filled symbols -- right component) for temperatures $T$=450 mK (left) and $T$= 10 K (right). Dashed curves for $T$= 450 mK are results of numerical modeling described in the text. Dashed curves for $T$= 10 K are fit results, data was fitted by equation $y=a(1-2Lx)$ with $L$=3.3.}\label{fig:widthleft}
	\end{figure}

	\begin{figure}[h]
		\includegraphics[width=\figwidth]{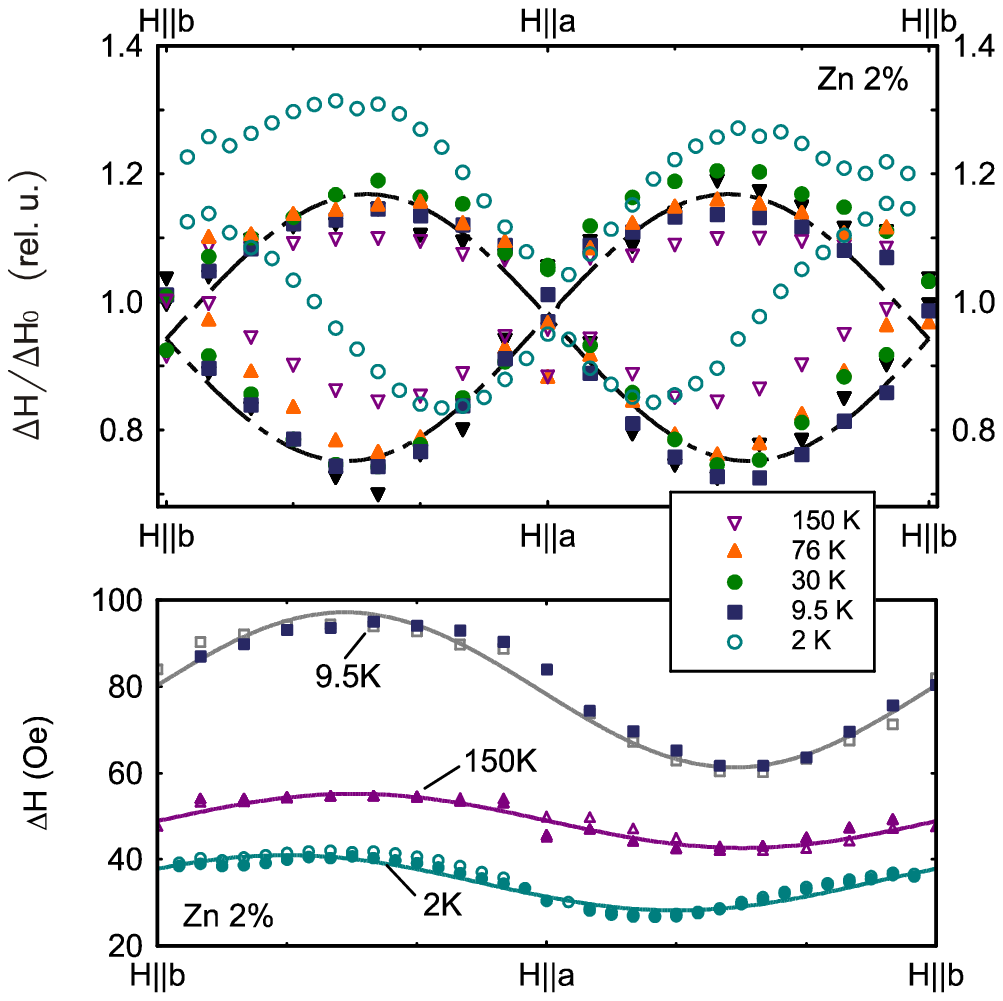}
		\caption{Upper panel: Angular dependencies of normed ESR linewidth for both components of ESR absorption lines for 2\% Zn-diluted DIMPY at different temperatures, $f=9.3$ GHz. All curves are normed by linewidth value at $H\parallel a$.  Different symbols correspond to different temperatures. Curves are guide to the eye. Lower panel: ESR linewidth at selected temperatures, $f=9.3$ GHz. Open and filled symbols correspond to different ESR line components, data for one of the components are translated to compensate for ladder nonequivalence. Curves are guide to the eye. }\label{fig:angular1}
	\end{figure}

	\section{Results and discussion.}
	\subsection{Reference results for pure DIMPY}
	ESR in pure DIMPY was reported in details earlier \cite{glazkov0, ozerov}. Examples of ESR absorption spectra are shown in the left panel of Fig. \ref{fig:lines}. Two components of the ESR absorption (at approximately 10.8 kOe and 12 kOe) correspond to the different ladders, which became inequivalent for the chosen field direction due to the $g$ - factor anisotropy. Intensities of these components are almost the same, small difference of the intensities can be ascribed to $g$-factor anisotropy and different relative polarizations of microwave field for non-equivalent ladders. Both absorption components lose intensity on cooling because of the energy gap and the ESR absorption in pure DIMPY practically vanishes below 500 mK. Additional splitting of the ESR components observed around 1 K is due to the zero-field splitting of the triplet sublevels, as discussed in details in Ref. [\onlinecite{glazkov0}].
	
	Analysis of the high-temperature ESR linewidth in pure DIMPY \cite{glazkov0} proved that the DM interaction is the main relaxation mechanism in DIMPY. This analysis allowed us to estimate the DM interaction strength ($D \simeq~0.3$K) and the direction of the DM vector; the DM interaction turns out to be  uniform along the legs and to be forbidden on the rungs of the ladder.
	
	\begin{figure}[h]
		\includegraphics[width=\figwidth]{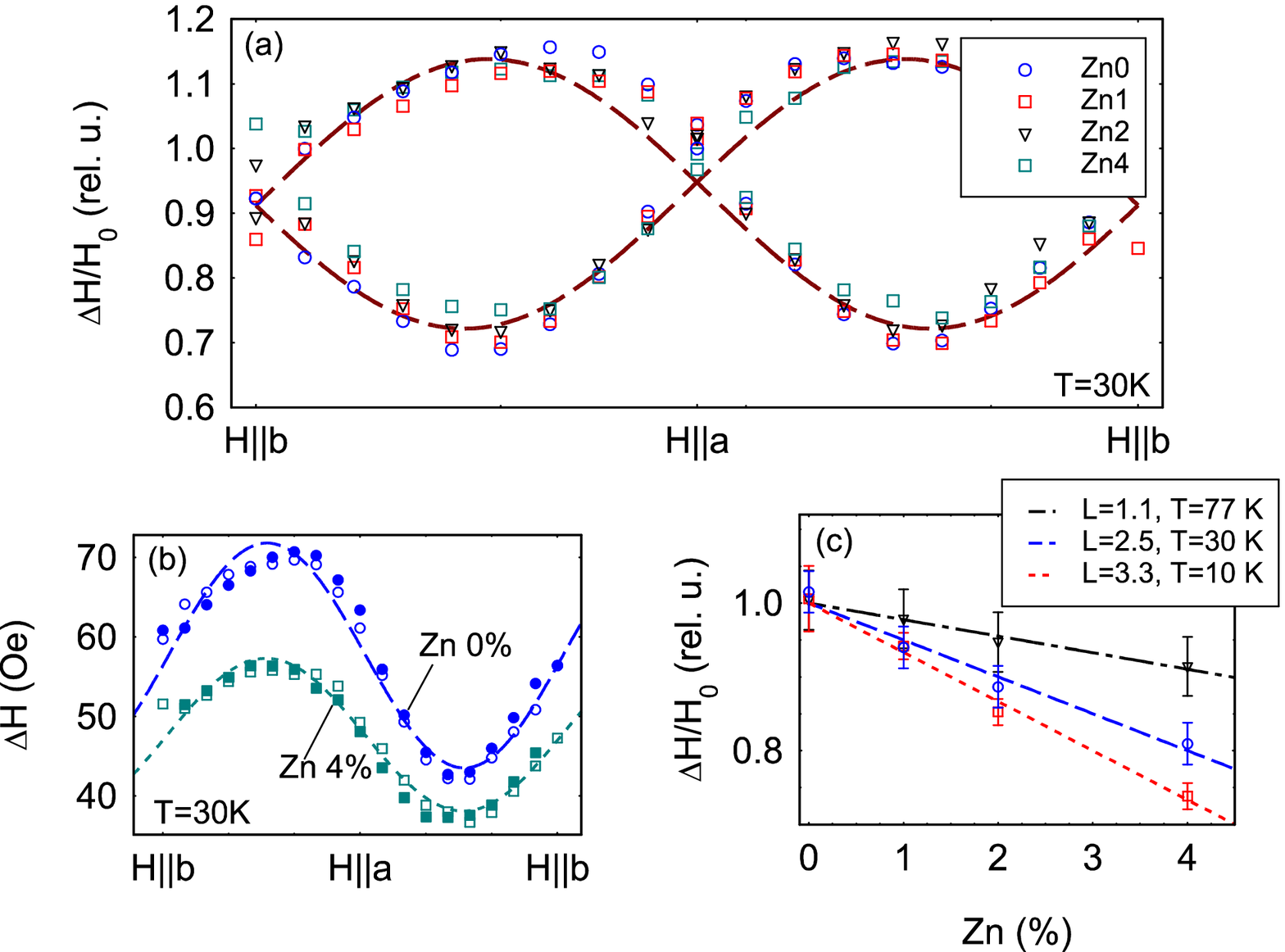}
		\caption{(a) Angular dependencies of normed ESR linewidth for $T=30$ K and different dilution level, $f=9.3$GHz.  All curves are normed by linewidth value at $H\parallel a$.  Different symbols correspond to different Zn concentrations. Curves are guide to the eye. (b) ESR linewidth for selected Zn concentration, $f=9.3$GHz. Open and filled symbols correspond to different ESR line components, data for one of the components are translated to compensate for ladder nonequivalence. Curves are guides to the eye. (c) Normed linewidth for $T$=10 K, 30 K, 77 K, and different zinc concentrations. Data are fitted by equation $y=a(1-2Lx)$ for each temperature.}\label{fig:angular2}
	\end{figure}
	
	\subsection{Temperature dependence of ESR absorption in Zn-diluted DIMPY.\label{sec:t-dependence}}
	Examples of the ESR absorption spectra for the diluted DIMPY are shown in Fig. \ref{fig:lines}.
	Contrary to the pure compound, the ESR signal in diluted DIMPY is intensive even at 450 mK, the ESR absorption intensity grows on cooling below 1 K in a Curie-like fashion. This indicates that some paramagnetic centers are formed on dilution. Intensity data (see insert to Fig. \ref{fig:lines}) enable us to highlight two regimes for diluted DIMPY: the impurity-dominated and the triplet-dominated. At the impurity-dominated regime (below approximately 2 K) the ESR absorption intensity increases on cooling, this increase can be phenomenologically fitted by the model of free spins $I \propto \tanh{(g{\mu}_{B}H/(2T))}$. At the triplet-dominated regime (above approximately 5 K) the ESR absorption intensity increases on heating as the population of the thermally activated triplets increases, this behavior can be qualitatively described by a 1D spin-gap magnet model $I \propto {\frac{1}{\sqrt{T}}}e^{{-\frac{\Delta}{T}}}$.
	
	The position of the ESR absorption does not change on cooling from the impurity-dominated to cluster-dominated regime and the $g$-factor value is the same for pure and diluted DIMPY. This similarity demonstrates that both for the triplet-dominated and impurity-dominated regimes, the ESR signal is due to the copper ions and there are no structural distortions in the magnet.
	
	These facts can be qualitatively interpreted starting from a strong-rung ladder model. The strong-rung spin ladder consists of weakly coupled dimers and substitution of one of the magnetic copper ions by a non-magnetic zinc leaves one unpaired spin one-half. This unpaired spin is responsible for the low-temperature magnetic response of the diluted system. Exchange coupling along the legs will spread this uncoupled spin along the ladder (on the scale of correlation length), leading to the formation of a multispin cluster (or ``spin island'', as it is called in Ref. [\onlinecite{schmidiger2}]) with the non-zero net magnetization.
	
	DIMPY is a strong-leg ladder, hence the magnetic correlation length is quite high ($\xi\sim 6...7$ interatomic distances) \cite{schmidiger2}. Consequently, clusters formed around different impurity ions could interact. This interaction was revealed through the low-temperature magnetization and the appearance of sub-gap states \cite{schmidiger2}. To check for the effects of cluster interaction we have measured temperature dependencies of the ESR absorption intensities for all available Zn concentrations. The ESR intensity is proportional to the spin magnetization, this allows us to scale our data to the known static susceptibility from Ref. [\onlinecite{schmidiger2}]. The resulting dependencies are shown in Fig. \ref{fig:intens}. The susceptibility curves obtained in the spin island model \cite{schmidiger2} are also shown in this figure. They deviate from our data below 1 K, which can be naturally ascribed to the saturation of the paramagnetic moment at low temperature since the ESR absorption occurs in the magnetic field around 10 kOe, while the magnetization curves were modeled at 100 Oe.

	We fit our scaled data in the model of free spins $S=1/2$ by:
	\begin{equation}
	I_{scaled}=\frac{M_z}{H}= x_{eff} \frac{g \mu_B N_A}{2H} {tanh\big(\frac{g\mu_B H}{2T}\big)}
	\end{equation}
	We will call the concentration of free spins $x_{eff}$ determined from this fit the apparent concentration of paramagnetic centers. There is a difference between the apparent concentration of paramagnetic centers and the nominal zinc concentration (see lower panel in Fig. \ref{fig:intens}). The apparent concentration has non-linear dependence from the real zinc concentration. This nonlinearity with negative curvature highlights the presence of the effective antiferromagnetic interaction between clusters.
	
	The nonlinearity of the apparent concentration can be simulated in the following simple model (Fig. \ref{fig:ladders}, lower panel). We assume that the average spin exponentially decreases away from the impurity $S(l)=(-1)^{l}S\exp(-l/\xi)$; here all distances are measured in interatomic distances along the leg of the spin ladder, $\xi=6$ is a correlation length for DIMPY \cite{schmidiger2}, and factor $(-1)^l$ explicitly takes into account antiferromagnetic spin-spin correlations. Overlapping of these exponential wings of the clusters results in the cluster-cluster coupling. The effective coupling energy for the clusters placed on the same leg of the ladder at distance $L_1$ can be estimated as $|E_{int}|=J_{leg} S^2 \exp(-L_1/\xi)$ and for the clusters placed on the different legs at distance $L_2$ as $|E_{int}|=J_{rung} S^2 \exp(-L_2/\xi)$.

	Cluster-cluster interaction tends to form coherent patterns of antiferromagnetic correlations. Depending on the parity of the distance between depleted rungs of the ladder this results in either parallel or antiparallel orientation of the clusters' spins (see Fig. \ref{fig:ladders}, lower panel, and Ref. [\onlinecite{schmidiger2}]). Thus, the effective cluster-cluster interaction can be ferro- or antiferromagnetic depending on cluster-cluster distance. Under the applied magnetic field; clusters tends to align their net magnetization along the magnetic field, in the case of the antiferromagnetic effective cluster-cluster coupling this leads to the competition between the Zeeman energy and $E_{int}$, which finally results in decrease of apparent concentration of paramagnetic centers. Energy cost of the ``flip'' of pair of clusters from $S_z=0$ state to $S_z=1$ state is $E_M=g\mu_B H$. We take that for the antiferromagnetic coupling (50\% of all cluster pairs) at $T\rightarrow 0$, the ``flip'' occurs if $E_M>E_{int}$, i.e., the apparent concentration is less than the real one by a fraction of strongly coupled closely positioned clusters. The threshold distances can be determined from the condition $E_{int}(L)=g\mu_B H$ which yields at 10 kOe field $L_1=7$ and $L_2=4$ for the clusters placed on the same leg and on different legs, correspondingly.
	
	The probability to find other impurities on the same leg at a distance less than $L_{1}$ or on another leg at a distance less than $L_{2}$ from the given impurity is:
	\begin{equation}
	w(L_1,L_2)=1 - (1-x)^{2L_{1}}\cdot (1-x)^{2L_{2}+1}
	\end{equation}
	Thus for the apparent concentration we obtain:
	\begin{eqnarray}
	x_{eff}&=&\frac{x}{2}+\frac{x}{2} \left(1-w(L_1,L_2)\right)=\nonumber\\
	&=&\frac{x}{2}({1+(1-x)^{2L_{1}}\cdot (1-x)^{2L_{2}+1}}))
	\end{eqnarray}
	For DIMPY parameters $x_{eff}=({x}/{2})({1+(1-x)^{23}})$, the result of this modeling is shown in the lower panel Fig. \ref{fig:intens}. The model curve fits the experimental data without any additional tuning parameters.

	\subsection{Linewidth in triplet-dominated regime and impurity-dominated regime.\label{sec:linewidth}}
	Another physical quantity that can be accessed in the ESR experiment is an ESR linewidth. It is determined by relaxation processes (essentially, the ESR linewidth is proportional to the inverse spin-relaxation time) and it is of interest how the dilution affects spin relaxation both in the impurity-dominated and the triplet-dominated regimes. The linewidth for pure DIMPY was discussed in details in Ref. [\onlinecite{glazkov0}]; this analysis demonstrated that the DM interaction determines spin-spin relaxation in pure compound. We have measured the temperature dependence of the ESR linewidth in diluted DIMPY from 0.45 K to 300 K (Fig. \ref{fig:widetemp}). To cover this temperature range, spectrometers operating at different microwave frequencies were used. We have checked that the ESR linewidth does not depend on frequency choice.

	Above 100 K, the ESR linewidth increases on heating presumably due to a spin-lattice relaxation. Between 5 K and 100 K (in the triplet-dominated regime), the linewidth temperature dependence for the diluted samples is similar to that for the pure compound with a maximum around 10 K. Temperature dependence of the ESR linewidth up to $T/J\simeq6$ is not unusual for low-dimensional systems, e. g., it was reported for 2D and 1D systems with DM interaction acting as a main source of the line broadening \cite{soos,fayzullin}. Maximum of the linewidth at 10 K ($T \sim J$) is due to crossover between different temperature limits: at higher temperatures   increase of the linewidth on cooling is due to the formation of short-range correlations in a paramagnet. At lower temperatures, a gap in the excitations spectrum (which is about 4 K) begins to affect excitations population: As temperature goes down excitations ($S=1$ triplons) freeze out, triplon-triplon interaction loses efficiency, triplon lifetime increases, and linewidth decreases on cooling. Microscopic theory of ESR in a strong-leg ladder \cite{furuya} also predicts the maximum of the ESR linewidth at $T \sim \Delta$.
	
	We have found (Fig. \ref{fig:widetemp}) that the ESR linewidth in diluted DIMPY in the triplet-dominated regime is smaller than in the pure compound, i.e., the spin relaxation time {\em increases} on dilution. In the impurity-dominated regime below 2 K, the ESR linewidth, in diluted DIMPY is higher than in the pure one, which is not surprising since the origin of the ESR signal in the pure and diluted compound is different in this temperature range.

	To understand the dilution effect on the ESR linewidth we have measured the temperature dependencies of the ESR linewidth for different Zn concentrations (Fig. \ref{fig:widthleft}). We have observed systematic narrowing of the ESR absorption line in the triplet-dominated regime (above 5 K) with Zn concentration growth (Fig. \ref{fig:widthleft}). It means that the spin relaxation time is the highest in the most diluted sample from the series ($x=$6\%).
	
	The spin relaxation is governed by anisotropic spin-spin interactions. To check that the relaxation mechanism remains unchanged with dilution we have measured the angular dependencies of the ESR linewidth for several temperatures from a wide temperature range (Fig. \ref{fig:angular1}) for all available concentrations of zinc (Fig. \ref{fig:angular2}).  We have noticed that the angular dependencies scale for different temperatures from 9.5 K to 77 K (Fig. \ref{fig:angular1}). The scaling of the angular dependencies with temperature means the presence of one dominant interaction involved in spin relaxation \cite{oshikawa}. Consequently, one dominant interaction is involved in the relaxation processes  for each measured concentration of Zn for a wide temperature range from 9.5 K to 77 K. Deviation from this scaling at 2 K and 150 K is not controversial because there are other regimes of relaxation at these temperatures.

	The ESR linewidth angular dependencies at fixed temperature for different zinc concentration (pure DIMPY included) also scale (Fig. \ref{fig:angular2}). This means that the same main anisotropic spin-spin interaction is responsible for the ESR linewidth in the triplet-dominated regime in both pure and diluted DIMPY and this interaction is the DM interaction. Moreover, the scaling of the ESR linewidth angular dependencies for different Zn concentrations proves that the direction of the DM vector remains unchanged with dilution.

	Thus, the linewidth decrease with dilution can be described as the effective DM vector shortening. The linewidth due to the DM interaction is proportional to $D^2/J$, here $J$ is some effective exchange coupling constant \cite{huber,glazkov0}. The gap in the excitation spectrum does not change significantly on dilution \cite{schmidiger2}, thus we expect that  the exchange coupling constant remains unchanged. The linewidth for the 6 \% diluted sample amounts at $T$=10 K to approximately 60 \% of the pure DIMPY linewidth at the same temperature, which means that the effective length of the DM vector is shortened by approximately 23 \% in this case: $|{\vec{D}(x=6 \%)}|\approx 0.77|{\vec{D}(x=0\%)}|$.

	The effective shortening of the DM vector can be modeled as total suppression of the DM interaction at the distance $\pm L$ from impurity (Fig. \ref{fig:ladders}, upper panel). This results in the decrease of the linewidth for different temperatures in the triplet-dominated regime:
	\begin{equation}
	\Delta H=\Delta H_0^{(HT)} (1-2 L x)
	\end{equation}
	The suppression length $L$ for $T$=10 K, 30 K, 77 K is equal to 3.3, 2.5, 1.1 interatomic distances correspondingly (Fig. \ref{fig:angular2}). Growth of the suppression length $L$ on cooling is probably linked to the increase of the magnetic correlation length. Value of $L$=3.3 is also used to fit the linewidth behavior with concentration growth at $T$=10 K at the higher microwave frequencies (Fig. \ref{fig:widthleft}). This fit uses the only scaling parameter $\Delta H_0^{(HT)}$ for the linewidth for the pure compound and well describes the experimental data.

	At temperatures below 2 K, we enter the impurity-dominated regime. Here concentration of the triplet excitations is negligible and the ESR response of diluted samples is governed by the multi-spin clusters (spin islands). Temperature dependence of ESR linewidth (Fig. \ref{fig:widthleft}) reaches saturation below approx. 1 K. We detected narrowing of the ESR line with dilution at $T<1$K (Fig. \ref{fig:widthleft}), which could be explained by the presence of the interaction between clusters: At the low dilution level, the spin relaxation due to some intra-cluster mechanism results in a ``seed'' linewidth of isolated cluster $\Delta H_0^{(LT)}$; at the higher dilution the effective exchange coupling between clusters $J_{eff}$ results in the narrowing of the ESR line through the exchange narrowing mechanism yielding the linewidth:
	\begin{equation}
	\Delta H=\frac{\left(\Delta H_0^{(LT)}\right)^2}
	{|{J_{eff}}|/(g\mu_B)}.
	\end{equation}
	As Zn concentration increases, the average distance between clusters decreases, the effective coupling grows, and the ESR linewidth decreases with the dilution.
	
	We calculated the average cluster-cluster coupling for different impurity concentrations by a ``brute-force'' numerical simulation of the ladder with randomly distributed defects for the known exchange integrals and correlation length of DIMPY (Fig. \ref{fig:ladders}, lower panel). This approach is similar to the model used to calculate the apparent concentration of paramagnetic centers in the previous subsection. The effective exchange integral for the clusters placed on the same leg of the ladder at a distance $l$ was estimated as $(-1)^{l}J_{leg}\exp(-l/\xi)$ and for the clusters placed on different legs as $(-1)^{l+1}J_{rung} \exp(-l/\xi)$. The absolute values of these couplings were averaged over randomly distributed defects. The result of this simulation was later used to fit the experimentally determined concentration dependence of the low-temperature ($T=0.45$K) linewidth, the only fitting parameter is a ``seed'' linewidth $\Delta H_0^{(LT)}$. Fit results are shown in Fig. \ref{fig:widthleft}. Best fit corresponds to the ``seed'' linewidth of the isolated cluster approximately equal to 490 Oe and 700 Oe for two absorption components.
	
	Additionally we can note here that we have not observed splitting of the ESR absorption line in diluted DIMPY around 1 K (see Fig. \ref{fig:lines}) even for the less diluted system. This is not surprising since the fine structure of the ESR line observed in the pure DIMPY at low temperatures is the consequence of zero-field splitting of the triplet excitations sublevels, while in diluted samples the ESR response of triplet excitations is overwhelmed by the clusters' response.
	
	\section{Conclusions.}
	The ESR response of the diamagnetically diluted strong-leg spin ladder \dimpy{} was studied from 450 mK to 300 K. At low temperatures (below 2 K), the ESR response is dominated by multi-spin clusters. The concentration dependence of the ESR intensity and linewidth provides evidences of the cluster-cluster interaction in agreement with Ref. [\onlinecite{schmidiger2}]. At higher temperatures (above 5 K), the ESR response is dominated by the collective triplet excitations. Analysis of the angular and temperature dependencies of the ESR linewidth proved that the spin relaxation mechanism is the same for pure and diluted samples and is governed by the DM interaction. Dilution does not change the direction of the DM vector, but shortens its effective length, which can be explained as the total suppression of DM interaction in the nearest vicinity (approximately $\pm$ 2 interatomic distances) of the defect.
	
	\acknowledgements
	We thank Dr. David Schmidiger (previously ETHZ) for his contributions
	in the early stages of this project and sample synthesis.
	We thank Dr. A.B. Drovosekov for help with measurements above 77 K.
	We also thank Prof. A. I. Smirnov, Prof. L. E. Svistov and Dr. T. A. Soldatov for supporting discussions.

	The work was supported by Russian Science Foundation Grant No. 17-12-01505 (experiments at Kapitza Institute at $T<77$ K), Russian Foundation for Basic Research 19-02-00194a (experiments at Kapitza Institute at $T>77$ K).
	Work in HSE was supported by Program of fundamental studies of HSE (data modeling and partial support of X-band measurements). Work in HZDR was supported by Deutsche Forschungsgemeinschaft
	(project ZV 6/2-2) and by the HLD at HZDR, member of the European Magnetic
	Field Laboratory (EMFL). Work at ETHZ was supported by
	the Swiss National Science Foundation, Division 2.


\begin{thebibliography}{10}
		\bibitem{ward} S. Ward, P. Bouillot, H. Ryll, K. Kiefer, K. W. Kr\"{a}mer, Ch. R\"{u}egg, C. Kollath and T. Giamarchi, J. Phys. Condens. Matter \textbf{25}, 014004 (2013).	
		
		\bibitem{ruegg} Ch. R\"{u}egg, K. Kiefer, B. Thielemann, D. F. McMorrow, V. Zapf, B. Normand, M. B. Zvonarev, P. Bouillot, C. Kollath, T. Giamarchi, S. Capponi, D. Poilblanc, D. Biner, and K.W. Kr\"{a}mer, Phys. Rev. Lett. \textbf{101}, 247202 (2008).
		
		\bibitem{hong} Tao Hong, Y. H. Kim, C. Hotta, Y. Takano, G. Tremelling, M. M. Turnbull, C. P. Landee, H.-J. Kang, N. B. Christensen, K. Lefmann, K. P. Schmidt, G. S. Uhrig, and C. Broholm, Phys. Rev. Lett. \textbf{105}, 137207 (2010).
		
		\bibitem{povarov} K. Yu. Povarov, D. Schmidiger, N. Reynolds, R. Bewley, and A. Zheludev, Phys. Rev. B, \textbf{91}, 020406(R) (2015).
		
		\bibitem{tsvelik} T. Giamarchi, A. M. Tsvelik, Phys. Rev. B, \textbf{59}, 11398 (1999).
		
		\bibitem{zapf} V. Zapf, M. Jaime, C. D. Batista, Rev. Mod. Phys. \textbf{86}, 563 (2014).
		
		\bibitem{giamarchi} T. Giamarchi, C. R\"{u}egg, O. Tshernyshov, Nature, \textbf{423}, 62, (2003).
		
		\bibitem{uehara} M. Uehara, T. Nagata, J. Akimitsu, H. Takahashi, N.
		Mori, and K. Kinoshita, J. Phys. Soc. Jpn. \textbf{65}, 2764 (1996).
		
		\bibitem{sigrist} M. Sigrist, T. M. Rice, and F. C. Zhang, Phys. Rev. B \textbf{49}, 12058 (1994).
		
		\bibitem{dagotto} E. Dagotto and T. M. Rice, Science \textbf{271}, 618 (1996).
		
		\bibitem{schmidiger} D. Schmidiger, S. M\"{u}hlbauer, A. Zheludev, P. Bouillot, T. Giamarchi, C. Kollath, G. Ehlers, A. M. Tsvelik, Phys. Rev. B \textbf{88}, 094411 (2013).
		
		\bibitem{white} J. L. White, C. Lee, \"O. G\"unaydin-\ifmmode \mbox{\c{S}}\else \c{S}\fi{}en, L. C. Tung, H. M. Christen, Y. J. Wang, M. M. Turnbull, C. P. Landee, R. D. McDonald, S. A. Crooker, J. Singleton, M.-H. Whangbo, and J. L. Musfeldt, Phys. Rev. B \textbf{81}, 052407 (2010).
		
		\bibitem{chaboussant} G. Chaboussant, Y. Fagot-Revurat, M. H. Julien, M. E. Hanson, C. Berthier, M. Horvatic, L. P. L\'evy, and O. Piovesana, Phys. Rev. Lett, \textbf{80}, 2713 (1998).
		
		\bibitem{shapiro} A. Shapiro, C. P. Landeee, M. M. Turnbull, J. Jornet, M. Deumal, J. J. Novoa, M. A. Robb, and W. Lewis, J. Am. Chem. Soc. \textbf{129}, 952 (2007).
		
		\bibitem{schmidiger0} D. Schmidiger, P. Bouillot, S. M\"{u}hlbauer, S. Gvasaliya, C. Kollath, T. Giamarchi, and A. Zheludev, Phys. Rev. Lett. \textbf{108}, 167201 (2012).
		
		\bibitem{schmidiger1} D. Schmidiger, P. Bouillot, T. Guidi, R. Bewley, C. Kollath, T. Giamarchi, A. Zheludev, Phys. Rev. Lett. \textbf{111}, 107202 (2013).
		
		\bibitem{glazkov0} V. N. Glazkov, M. Fayzullin, Yu. Krasnikova, G. Skoblin, D. Schmidiger, S. M\"{u}hlbauer, and A. Zheludev, Phys. Rev. B \textbf{92}, 184403 (2015).
		
		\bibitem{ozerov} M. Ozerov, M. Maksymenko, J. Wosnitza, A. Honecker, C. P. Landee, M. M. Turnbull, S. C. Furuya, T. Giamarchi, and S. A. Zvyagin, Phys Rev. B \textbf{92}, 241113(R) (2015).
		
		\bibitem{sandvik} A. W. Sandvik,  E. Dagotto, and D. J. Scalapino, Phys. Rev. B \textbf{56}, 11701 (1997).
		
		\bibitem{poilblanc} N. Laflorencie and D. Poilblanc, Phys. Rev. Lett \textbf{90}, 157202 (2003).
		
		\bibitem{yasuda} C. Yasuda, S. Todo, M. Matsumoto, and H. Takayama, Phys. Rev. B \textbf{64}, 092405 (2001).
		
		\bibitem{wessel} S. Wessel, B. Normand, M. Sigrist, and S. Haas, Phys. Rev. Lett \textbf{86}, 1086 (2001).
		
		\bibitem{fukuyama} H. Fukuyama, T. Tanimoto, and M. Saito, J. Phys. Soc. Jpn. \textbf{65}, 1182 (1996).
		
		\bibitem{glazkov1} V. N. Glazkov, Yu. V. Krasnikova, D. H\"{u}vonen, and A. Zheludev, J. Phys. Condens. Matter \textbf{28}, 206003 (2016).
		
		\bibitem{martins} G. B. Martins, M. Laukamp, J. Riera, and E. Dagotto, Phys. Rev. Lett \textbf{78}, 3563 (1997).
		
		\bibitem{mikeska} H.-J. Mikeska, U. Neugebauer, U. Schollw\"{o}ck, Phys. Rev. B \textbf{55}, 2955 (1997).
		
		
		
		\bibitem{schmidiger2} D. Schmidiger, K. Yu. Povarov, S. Galeski, N. Reynolds, R. Bewley, T. Guidi, J. Ollivier, A. Zheludev, Phys. Rev. Lett. \textbf{116}, 257203 (2016).
		
		\bibitem{soos} Z. G. Soos, K. T. McGregor, T. T. P. Cheung, and A. J. Silverstein, Phys. Rev. B \textbf{16} 3036 (1977).
		
		\bibitem{fayzullin} M. A. Fayzullin, R. M. Eremina, M. V. Eremin, A. Dittl, N. van Well, F. Ritter, W. Assmus, J. Deisenhofer, H.-A. Krug von Nidda, and A. Loidl, Phys. Rev. B \textbf{88}, 174421 (2013).
		
		\bibitem{furuya} S. C. Furuya and M. Sato, J. Phys. Soc. Jpn. \textbf{84}, 033704 (2015).
		
		
		\bibitem{oshikawa} M. Oshikawa and I. Affleck, Phys. Rev. B, \textbf{65}, 134410 (2002).
		
		\bibitem{huber} D. L. Huber, G. Alejandro, A. Caneiro, M. T. Causa, F. Prado, M. Tovar, and S. B. Oseroff, Phys. Rev. B \textbf{60}, 12155 (1999).
		
	\end{thebibliography}
\end{document}